**Mechanisms for impulsive energy dissipation and small scale effects in micro-granular media**

Jonathan Bunyan, Alexander F. Vakakis and Sameh Tawfick

*University of Illinois at Urbana-Champaign*

### Abstract

We study impulse response in 1-D homogeneous micro-granular chains on a linear elastic substrate. Micro-granular interactions are analytically described by the Schwarz contact model which includes nonlinear compressive as well as snap-to/from-contact adhesive effects forming a hysteretic loop in the force deformation relationship. We observe complex transient dynamics, including disintegration of solitary pulses, local clustering and low- to high-frequency energy transfers resulting in enhanced energy dissipation. We study in detail the underlying dynamics of cluster formation in the impulsively loaded medium, and relate enhanced energy dissipation to the rate of cluster formation. These unusual and interesting dynamical phenomena are shown to be robust over a range of physically feasible conditions, and are solely scale effects, since they are attributed to surface forces, which have no effect at the macro-scale. We establish a universal relation between the re-clustering rate and the effective damping in these systems. Our findings demonstrate that scale effects generating new nonlinear features can drastically affect the dynamics and acoustics of micro-granular materials.

# I.    INTRODUCTION

The dynamics of ordered granular chains in the macro-scale is non-linear, and posses a rich plethora of tunable dynamic mechanical behaviors. In particular, the impulsive responses of un-compressed macro-granular chains composed of spherical elastic beads in contact are highly nonlinear, in fact, being characterized as 'sonic vacua' [1] due to absence of linear acoustics and vanishing speed of sound as defined in classical linear acoustics. In the macro-scale these highly discontinuous passive media support Nesterenko solitary waves [1] and have been applied to stress wave trapping and arrest [2,3], frequency banding and acoustic filtering [4], dynamics and acoustics that are tunable with energy [1,5], nonlinear acoustic lensing [6,7], and nonlinear stress wave tailoring [8-10]. The strongly nonlinear nature of macro-scale Hertzian interactions stems from two types of nonlinearities: first, the non-linearizable Hertzian interaction law between elastic particles in contact [11], and second, the possibility of loss of contact in the absence of compressive forces and ensuing collisions between particles rendering the constitutive law non-smooth. Both these non-linear effects pose significant challenges to the analytical study of the dynamics and acoustics of these media, and prevent the application of commonly used techniques based on linearized approaches, homogenization, and classic asymptotic analysis.

Considering granular media in the micro- or nano scale, it is expected that the physics governing their dynamics and acoustics will be altered, which to our knowledge, has not been studied previously. At these small scales, in addition to Hertzian-like interactions which govern the compressive regime, non-negligible attractive surface forces, known as van der Waals forces, are known to change the overall force-deformation contact law and introduce additional strong nonlinearities, including snapping to and away-from contact and adhesion hysteresis [12-14]. Yet, the study of the nonlinear response of ordered granular chains in the micro-scale is still in its infancy. In their pioneering work, Boechler and co-workers observed "avoided crossings" in the dispersion relations between local microsphere resonances and substrate Rayleigh waves in micro-granular lattices. This was accomplished by studying the contact resonances of a 2-D hexagonal close-packed monolayer of soft polystyrene microparticles interacting with long wavelength surface acoustic waves on a silicon substrate excited through laser-induced transient grating [10,15]. However, in that work the observed behavior was governed by the particles oscillating as individual resonators on an elastic substrate, rather than in terms of waves propagating through the micro-granular medium and governed by interactions between neighboring microparticles. Accordingly, loss of contact between adjacent microparticles was not considered.

In this study we consider longitudinal wave propagation in an impulsively forced finite 1-D micro-granular 'stiff' chain composed of linearly elastic spherical beads laying on a 'soft' tunable elastic substrate, and take into account the possibility of loss of contact between neighboring beads, as well as adhesion induced hysteresis due the micro-scale effect. We find that the classical Nesterenko solitary pulse [1] propagating with no dispersion seen in the macro-scale, quickly disintegrates in the micro-scale, and that phenomena such as local clustering and energy transfers from low-to-high frequencies emerge and drastically alter the effective properties of the micro-granular medium compared to equivalent macro-granular ones. We study in detail the underlying nonlinear dynamics and their relationship to the changes in effective material properties of this medium, as well as the robustness and tunability with energy (impulse intensity) of these phenomena. Due to the strongly nonlinear nature of this system and our interest in the long-time dynamics, our study is performed computationally.

## II.     MODEL AND GOVERNING EQUATIONS

We consider the transient impulsive responses of one-dimensional micro-granular chains, to study the nonlinear dynamics and observe changes in behavior compared to an equivalent system in the macro-scale. The micro-granular chain [cf. Fig 1(a)] is composed of $N$ identical spherical ceramic (Alumina) beads of $2.5\,\mu m$ diameter subject to impulse excitation. The chain is bounded by fixed spherical beads of the same diameter. An ideal impulse is applied to the first bead and each bead is modeled as a point mass interacting nonlinearly with its neighbors [1], supported by an elastic substrate modeled as an elastic foundation. We note that this is considered to represent the effect of the chain's interaction with any supporting or confining surface. We represent the foundation stiffness by a tensioned thin wire, which, because of its geometry and kinematics, has both linear and cubic terms, and could enable in the future the tuning of the support stiffness. The inter-bead interaction is modeled by the Schwarz contact model, which considers the influence of short range as well as long range attractive forces both inside and outside the actual contact area [12]. The model captures the popular JKR (soft-high adhesion interactions) [13] and DMT (hard interactions) [14] theories, and describes the hysteretic interactions in the intermediate contact regime through an analytical expression relating the force between micro-beads, $F_\mu$, to the deformation of the center distance between adjacent beads, $\delta$, as depicted in Fig.1(b), and defined through Eqs. (1) and (2) below,

$$a = (R/K)^{1/3}(\sqrt{3F_c + 6\pi Rw} \pm \sqrt{F_\mu - F_c})^{2/3}, \tag{1}$$

$$\delta = \frac{a^2}{R} - 4\sqrt{\frac{\pi a}{3K}\left(\frac{F_c}{\pi R} + 2w\right)}, \tag{2}$$

where $a$ is the contact radius of the interacting beads, $K = (2/3)E(1-\nu^2)^{-1}$ the mechanical modulus, $R = R_1 R_2 / (R_1 + R_2)$ the effective local radius ($R_1$ and $R_2$ are the radii of the interacting beads), $E = 338 GPa$ the Young's modulus, $\nu = 0.21$ the Poisson's ratio, $w = 0.146 Jm^{-2}$ the work of adhesion and $F_c$ the critical force that acts as a transition parameter varying between $F_c = -(3/2)\pi Rw$ (JKR limit) and $F_c = -2\pi Rw$ (DMT limit). The transition parameter $F_c$ is related to the more common Maugis transition parameter $\lambda$ through the Carpick-Ogletree-Salmeron empirical approximation, $\lambda = -0.924 \ln(1 - 1.02\sqrt{-3 - 6\pi Rw / F_c})$, where $\lambda = 2.06 / D_0 (Rw^2 / (\pi K^2))^{1/3}$ is calculated to be 0.64 for the chosen material properties and geometry [12,16,17]. The work of adhesion is theoretically calculated as $w = A_{12} / 12\pi D_0^2$ where $D_0$ is the interfacial contact separation and $A_{12} = 15(10^{-20} J)$ is the non-retarded Hamaker constant for two alumina surfaces interacting in a vacuum (inert air) at room temperature. The interfacial contact separation is generally accepted to be $D_0 = 0.165nm$ for a variety of ordinary common materials, and this value has been experimentally proven to be accurate within 10-20% [18].

The hysteretic nature of this interaction is shown in Fig.1(b). According to this model, there is no force when beads are approaching one another until contact has been established at $\delta = 0$; however, once contact has been established limited tensile loads can be supported during separation for $\delta < 0$ since the beads extend the effective range of contact by deforming when pulled apart. More precisely, in the tensile regime surface adhesion forces deform the particles which eventually snap apart at the minima of $\delta(a)$ from Eq. (2). In the compressive regime ($\delta > 0$) the force-deformation relationship is qualitatively similar (but not identical) to Hertzian contact described by $F_H(\delta) = K\sqrt{R}\delta_+^{3/2}$, with subscript (+) indicating that $F_H = 0$ when contact is lost for $\delta < 0$ [Fig. 1(b)]. It should be noted that in the limit of $F_c = -2\pi Rw$ the hysteresis vanishes and the DMT model is recovered. We set the initial conditions such that the beads are in contact without being deformed; accordingly, due to attractive forces the chain is considered to be in an initial state of pre-tension ($F_\mu < 0$) as shown in Fig. 1. In addition to hysteresis, material (grain boundary) dissipation is also included and modeled as weak unilateral linear viscous damping between beads with damping coefficient $C_+ = 0.1\mu Nsm^{-1}$ [19] where subscript (+) indicates that the damping coefficient is non-zero only in the contact regime [Fig. 1(a)]. Note that viscous damping has been proven to provide a good approximation to granular dissipative dynamics [20-22]. We emphasize again that the natural equilibria of the beads in an unsupported micro-granular chain is a state of compression due to the attractive forces; this initial state of compression, however, can be counterbalanced by the forces provided by the elastic substrates of the beads. Hence, in this model the fixed boundary conditions and elastic substrate keeps the beads in an un-deformed initial state, but as a consequence, finite pre-stored potential energy is 'deposited' in the elastic foundation of the microgranular chain. As described above, the forces provided by the elastic substrate are due to deformations of taut elastic micro-wires supporting each of the micro-beads, providing a restoring force equal to,

$$F_s(x_i) \approx \left(\frac{2T_s}{L_s}\right)x_i + \left(\frac{E_s A_s - T_s}{L_s^3}\right)x_i^3, \quad i = 1, ..., N, \tag{3}$$

where $F_s$ is the force acting on the beads from the wires [23], $x_i$ the bead displacement [cf. Fig. 1(a)], $T_s$, $A_s$ and $L_s$ the wire pretension, cross section and natural length, respectively, and $E_s = 400\,GPa$ the Young's modulus of the wire. In this study, we focus on the range of wire tension where stiffness is dominated by the linear coefficient $K_1 = 2T_s / L_s$ and the cubic term is safely neglected. Moreover, the grounding is chosen to be much 'softer' compared to the 'stiffer' inter bead dynamics described in Eq. (1) and (2) so that a corresponding time-scale separation between the stiff granular dynamics and the soft elastic foundation forces occurs. The governing nonlinear equations of motion of the micro-granular chain are then expressed as,

$$
\begin{aligned}
m_i \ddot{x}_i &= F_\mu(x_{i-1} - x_i + \delta_0) - F_\mu(x_i - x_{i+1} + \delta_0) - K_1 x_i \\
&+ (\dot{x}_{i-1} - \dot{x}_i)C_+ - (\dot{x}_i - \dot{x}_{i+1})C_+, \quad i = 1, ..., N.
\end{aligned}
\tag{4}
$$

with $x_0 = x_{N+1} \equiv 0$. In Eq. (4) $m_i$ is the mass of the $i-th$ bead, and the parameter $\delta_0$ is the initial elastic overlap between beads and set to $\delta_0 = 0$. The equations of motion are written in non-dimensional form using the transformations, $x = p\hat{x}$, $F_\mu(\delta) = \eta\hat{F}_\mu(\delta / p)$ and $\tau = qt$. Applying these transformations to the original equation we obtain,

$$
\begin{aligned}
m_i p q^2 \hat{x}_i'' &= \eta\hat{F}_\mu(\hat{x}_{i-1} - \hat{x}_i) - \eta\hat{F}_\mu(\hat{x}_i - \hat{x}_{i+1}) - K_1 p\hat{x}_i \\
&+ pq(\hat{x}_{i-1}' - \hat{x}_i')C_+ - pq(\hat{x}_i' - \hat{x}_{i+1}')C_+, \quad i = 1, ..., N,
\end{aligned}
\tag{5}
$$

where prime denotes differentiation with respect to rescaled time $\tau$. Dividing by $m_i p q^2$ and setting $q^2 = \eta / (m_i p)$ we obtain the non-dimensional equation of motion for the $i - th$ bead:

$$\hat{x}_i'' = \hat{F}_\mu(\hat{x}_{i-1} - \hat{x}_i) - \hat{F}_\mu(\hat{x}_i - \hat{x}_{i+1}) - \left(\frac{K_1 p}{\eta}\right)\hat{x}_i$$
$$+ \left(\frac{pqC_+}{\eta}\right)(\hat{x}_{i-1}' - \hat{x}_i') - \left(\frac{pqC_+}{\eta}\right)(\hat{x}_i' - \hat{x}_{i+1}'), \quad i = 1,...,N, \tag{6}$$

where $p = \left(\pi^2 R^2 w^2 K^{-2}\right)^{1/2}$ and $\eta = \pi R w$ represent the characteristic length and force scales for the inter-bead interaction model. The equations of motion subject to initial conditions are integrated numerically in a sequence of small piecewise continuous segments each terminating at jump discontinuities in inter-bead forces due to changes in the contact state.

## III.     RESULTS AND MICRO SCALE EFFECTS

In Fig. 2(a, b) we depict the spatiotemporal evolution of the instantaneous kinetic energies of impulsively excited macro- and micro-granular chains with $N = 20$ beads. As a reference, we compare the impulse response of the micro-beads to an equivalent macro-scale granular chain. We use normalized time scales by dividing the time variable in each of the two simulations by the time needed for the primary Nesterenko solitary pulse to travel twice from one boundary to the other in each case; moreover, in each case the applied impulse energy is scaled by matching the corresponding speeds of the primary Nesterenko solitary pulses in the macro- and micro-systems. The wire length, $L_s$, and diameter, $d_s$, are scaled geometrically with the bead size, $d$, whereas the initial substrate pretension is adjusted so that $K_1$ is non-dimensionally equivalent for the systems to ensure that it introduces same-scale soft dynamics relative to the stiffer inter-bead dynamics; the damping coefficient $C_+$ is also scaled similarly. For reference the stiffness and damping ratios for the two systems are tabulated in Table I, together with the corresponding normalized coefficients used in Eq. (6). The applied wire tensions $T_s$ are tabulated in Table II, and for reference, the maximum uniaxial tensile stress of the wire material is shown.

*(Figure 2)*

TABLE I. Normalization factors and coefficients $K_1$ and $C_+$ for the two granular chains.

| $d$ | $(\pi R^2 w K^2)^{1/3}$ $\left[kNm^{-1}\right]$ | $m_i^{1/2}(\pi R^2 w K^2)^{1/6}$ $\left[Nsm^{-1}\right]$ | $K_1$ $\left[Nm^{-1}\right]$ | $C_+$ $\left[Nsm^{-1}\right]$ |
|---|---|---|---|---|
| $2.5\mu m$ | 1.776 | $7.53\mu$ | 21.5 | $0.100\mu$ |
| $2.5mm$ | 177.584 | 2.38 | 2150 | 0.0316 |

TABLE II. Properties of the taut wires required for the desired substrate stiffness $K_1$ and maximum permissible uniaxial tensile stress in the wire material

| $d$ | $L_s$ | $d_s$ | $K_1$ | $T_s$ | $\sigma_y$ |
|---|---|---|---|---|---|
| $2.5\mu m$ | $6.25\mu m$ | $1\mu m$ | $21.5Nm^{-1}$ | $67.2\mu N$ | 2.19 MPa |
| $2.5mm$ | $6.25mm$ | $1mm$ | $2150Nm^{-1}$ | $6.72N$ | 0.219 MPa |

To be able to visually observe the dynamics of the granular chains for an extended period of time and study the impulsive energy decay, the kinetic energy is normalized at each time step by the maximum kinetic energy of any bead at that time instant. In the micro-granular chain the initial solitary Nesterenko solitary pulse quickly disintegrates due to the micro-scale attractive forces, and this is accompanied by bead cluster formation, unusual fragmentation of the chain, and spatial localization of the kinetic energy as shown in Fig. 2(b). Cluster boundaries or "fragmentation" (breakage) in the chain defined by loss of contact between micro-beads is represented by solid horizontal lines in Fig. 2(b). Bead clustering is the result of instantaneous dynamic balance of near-field attractive, elastic, and substrate forces, so cluster disintegration and reformation continuously occurs until the beads have insufficient energy to escape from their respective clusters. The elastic substrate, in addition to introducing slow-time effects, tunes the formation and sizes of the clusters, including those not directly connected to the fixed boundaries. These effects, however, are not observed in the macro-granular chain [Fig. 2(a)] where there is sustained propagation of the Nesterenko solitary pulse, with small dispersion appearing at later times due to the elastic substrate and dissipation due to viscoelastic effects. In the simulations for the macro-granular chain the Schwarz contact model $F_\mu(\delta)$ is replaced by the Hertzian contact model $F_H(\delta)$ in Eq. (4), and the same non-dimensionalization is applied as in the micro-granular chain. The use of the simpler Hertzian model to represent the macro scale case is justified in that for macro-beads both, the Schwarz and Hertzian models produce similar results since the attractive forces between beads are negligible compared to the compressive ones. Comparing the results of Figs. 2(a) and 2(b) we conclude that the observed dynamic behavior in the micro-granular chain is caused by the adhesive nature of the micro-beads which allows for *a state of self-stress*. The balance of forces between the soft grounding stiffness, the elastic contact forces and the near-field attractive forces between beads leads to the formation of clusters of micro-beads which can temporarily localize kinetic energy. For example, the plot of Fig. 2(b) shows kinetic energy being localized in beads 11-20 following the disintegration of the Nesterenko solitary pulse.

The formation of temporary clusters in the micro-granular chain is a scale-effect phenomenon, and alters the system response compared to the macroscale. The contact model indicates that there can be no energy transfer between neighboring clusters until cluster disintegration and re-clustering takes place, causing spatial redistribution of energy in the newly formed clusters. Fig. 2(c) shows the evolution of the total kinetic energies in the two chains normalized by the impulse energy. As it becomes apparent from Fig. 2(b, c), cluster disintegration and reformation rate is related to the kinetic energy evolution, and continues until the beads within clusters do not have sufficient energy to escape and join neighboring clusters. There is notably faster dissipation of kinetic energy in the micro-granular chain, indicating an increase in effective dissipation. In the case of micro-beads the normalized kinetic energy is observed to increase beyond its initial value as stored strain energy is released and converted into kinetic energy during clustering events. Interestingly, even with this influx of energy into the system, the total kinetic energy in the system decays faster compared to the macro-bead system.

To explain this result, the effective dissipation in the system is assessed quantitatively. Dissipation measures in linear materials, such as $\tan \Delta$, have been defined to study steady state inherent dissipation under harmonic excitation, where $\Delta$ is the phase between stress and strain in steady state harmonic vibration [24]. Clearly, such linear (or linearized) measures are inapplicable to the strongly nonlinear transient dynamics of the granular chains under consideration. Hence, we propose the following alternative effective damping measure based on direct processing of bead transient responses. To this end, for a given granular chain response and at each time instant we define an 'equivalent' single degree of freedom system with effective mass $m_{eff} = Nm$, deformation $x_{eff}(t)$, and time varying stiffness $K_{eff}(t)$ and damping $C_{eff}(t)$, which up to that time instant has accumulatively dissipated an equal amount of energy as in the granular chain under study. It follows that the energy dissipated accumulatively by the equivalent model is given by:

$$E_{d,eff}(t) = \int_{0}^{t} C_{eff} \dot{x}_{eff}^2 dt. \tag{7}$$

Realizing that $2T_{eff} = m_{eff} \dot{x}_{eff}^2$, the rate of change of dissipated energy is expressed as,

$$\dot{E}_{d,eff}(t) = 2C_{eff}(t)\frac{T_{eff}(t)}{m_{eff}}, \qquad (8)$$

where $T_{eff}$ is the instantaneous effective kinetic energy of the single degree of freedom system. For the $N$ bead system, the accumulated dissipated energy up to the given time instant is given by $E_d(t) = E_0 - T(t) - V(t)$, where $E_0$ is the initial energy, and $T(t)$, $V(t)$ the instantaneous kinetic and potential energies. By enforcing that the instantaneous kinetic energy and energy dissipated accumulatively are instantaneously equal for both systems, and implementing numerical averaging to smoothen the data, the time varying effective damping coefficient can be expressed as,

$$C_{eff}(t) = \frac{mN}{2}\frac{d\langle E_d(t)\rangle}{dt}\frac{1}{\langle\langle T(t)\rangle\rangle}, \qquad (9)$$

where $\langle\bullet\rangle$ denotes a smooth cubic spline fit, and $\langle\langle\bullet\rangle\rangle$ the mean of the cubic spline interpolants of the envelopes of the minima and maxima of the instantaneous kinetic energy in time. Fig. 2(d) depicts the normalized effective damping coefficient $\overline{C}_{eff}(t)$ [calculated through normalizing Eq. (9) by the linear viscous damping coefficient $C_+$] for the impulsively excited micro- and macro-granular chains. Comparing $\overline{C}_{eff}(t)$ for the micro granular chain and the clustering formation in Fig. 2(b), we note that increases in effective damping correspond to clustering and re-clustering events following the disintegration of the initial Nesterenko solitary pulse. We observe that a consistent feature in the considered cases is that the increase in effective damping coincides with for the rapid decrease in kinetic energy in Fig.2(c). Importantly, the final stored energy of the system is less than the initial stored energy, indicating dissipation of *initial potential and kinetic energy* owing to the unique transient nonlinear dynamics. As shown below, these phenomena are robust and observed over a range of impulse energies and substrate stiffness.

## IV.    MECHANISMS OF ENERGY DISSIPATION IN MICROGRANULAR CHAINS

We now study in detail the different mechanisms responsible for the observed enhanced effective damping in the microscale, and correlate them to the transient nonlinear dynamics of the micro-granular chain. The salient feature of the micro-granular chain is the temporary clustering, cluster disintegration, and re-clustering of the micro-beads. In particular, re-clustering events give rise to relatively high frequency relative oscillations of the compressed beads inside early transient clusters, causing increased dissipation within the clusters due to viscoelastic interactions between beads. Moreover, effective damping is further enhanced by the hysteretic loop illustrated in Fig. 1(b) which becomes profound at the instants of cluster formation, escape and re-clustering. This explains the sudden jumps in the normalized energy and effective damping dissipation measures of Figs. 2(c, d). On the contrary, these two dissipative effects are absent in the macro-granular chain where impulsive energy is transferred by the (near-zero frequency) Nesterenko solitary pulse and no clustering occurs (due to the absence of attractive forces), so the corresponding effective dissipation measure is smaller and slowly varying. Below we discuss these dynamical phenomena in more detail.

## A.    Linearized dynamics within clusters

The formation, disintegration and re-formation of local clusters in the micro-granular chain of Fig. 2(b) are reconsidered in Fig. 3(a) with dashed boxes representing clusters defined by temporary loss of contact between micro-beads. The resulting energy localization at the clusters is clear in Fig. 3(a) where propagating pulses can be seen reflecting within cluster boundaries and energy is localized away from beads 1-10 after about 1.6 characteristic periods (a characteristic period is defined as twice the duration of the Nesterenko primary pulse). Given that beads within a cluster are in a condition of compression and that their oscillation amplitudes are sufficiently small, the dynamics inside a cluster can be studied by linearizing Eq. (6) around the natural equilibrium [denoted by the star in Fig. 1(b)] and solving the resulting eigenvalue problem. Previous results on finite homogeneous macro-granular chains showed that, typically, most of the energy of propagating pulses can be captured by in the lowest frequency in-phase mode (standing wave) [25]. In Fig. 3(b-e) we depict the wavelet transform spectra of the middle beads of transient clusters I through IV of Fig. 3(a). Dashed horizontal lines in the same plots indicate the first (in-phase) and second fundamental frequencies predicted analytically by linearization of the dynamics of the corresponding cluster, and superimposed to these are the wavelet transform spectra of the middle beads of each cluster.

The agreement between the wavelet spectra and the linearized predications for the fundamental modes of the clusters indicate that the dynamics within each transient cluster is almost linear due to the compression between micro-beads generated during cluster formation. As expected, the dynamics in higher energy clusters I, II and III deviate further from the linearized predicted modes than the dynamics in the low energy cluster IV. The slight difference in the low energy cluster IV is thought to be due to the neglected non-uniformity of strain in the elastic substrate. Fig. 4 shows the spatiotemporal variation of the amplitudes of relative velocities between neighboring beads after the impulse in the micro-granular chain has been applied. Note that within the persisting clusters II, III and IV the viscous damping mechanism is rather ineffective as it dissipates energy proportional to the small relative motion of beads in the in-phase mode; this agrees with the small values of the effective damping measure $\overline{C}_{eff}$ for at low energies in Fig. 2(d). On the contrary, a high re-clustering rate (leading to short-lived clusters) gives rise to relatively high relative velocities between neighboring beads and to corresponding high effective damping measures. Indeed, the time instants of high relative velocity amplitudes in Fig. 4 correlate to high re-clustering rate in Fig. 3(a) and high effective dissipation measure for the micro-granular chain in Fig. 2(d). We note also the relatively high-frequency bead oscillations within the clusters (compared with the near-zero frequency solitary pulse propagation in the macro-granular system), and the fact that the temporally linear dynamics within each cluster coexist with re-clustering phenomena occurring in other regions of the micro-granular chain. Similar linear analysis cannot be performed for highly transient clusters with lifetimes much shorter than the first fundamental period; only higher harmonics and inherently nonlinear sub-harmonics may be observed in these clusters and the effective particle approach does not apply.

_**(Figure 4)**_

## B.      Energy dissipation due to nonlinear re-clustering

In this section, we describe the cluster dynamics and its relation to the effective damping measure. Considering the plot of Fig. 4 we note that unlike the linear dynamics within persisting (long-lived) clusters, local increases in relative velocity occur close to 0.2 and 0.3 μs, and coincide with increase in density of re-clustering events as well as peaks in instantaneous kinetic energy [cf. Fig. 4 and 2(c)]. Hence, during high-rate transient re-clustering, both hysteretic and viscous damping mechanisms are active.  Accordingly, the increase in effective damping is a result of higher frequencies introduced through short duration transient re-clustering rather than long duration linearized dynamics within more stable clusters. Moreover, these transient high-frequency oscillations are amplified by the release of the pre-stored potential energy in the granular chain. While it can be argued that attractive micro-scale forces in effect amount to negative stiffness [note the snapping to and away from contact in Fig. 1(b)] and can introduce unstable states in a system [26-29], in this case they are stabilized by boundary and elastic substrate forces. Hence, clustering events represent instantaneous dynamic balance between these effects.

These results indicate that micro-scale granular materials, supported by elastic substrates against agglomeration, posses significantly higher energy dissipation and effective damping than their macro-scale counterparts owing to attractive surface forces. To quantitatively study this effect, the cumulative number of newly formed clusters is calculated, and its time derivative is plotted in Fig. 5 for the system of interest to represent the instantaneous re-clustering rate (See Appendix A for details). This measure of re-clustering rate correlates to the re-clustering density depicted in Fig. 2(b), and shows that high re-clustering rates are categorically related to peaks in effective damping in this system. This will be discussed further below in light of the important issue of robustness of these nonlinear phenomena for changes of the stiffness of the elastic substrate and the intensity of the applied impulse. We note that slightly negative re-clustering rates in Fig. 5 are artifacts of the smoothing algorithm, and, hence, are non-physical.

*(Figure 5)*

## V.  ROBUSTNESS FOR VARYING STIFFNESS AND IMPULSE ENERGIES

The nonlinear phenomena described in the previous section exist and are robust over a range of impulse energies and substrate stiffness. Fig. 6(a, b) depicts the spatiotemporal kinetic energy distribution and effective damping coefficients for four cases (see Appendix B for more test cases). The effective damping measure is consistently observed to increase rapidly following the disintegration of the initial Nesterenko solitary pulse and decrease once the clusters have become persistent. The previously discussed relationship between re-clustering rates and peaks in effective damping is observed to be robust even for delayed peaks in effective damping— for example, the lowest two impulse energies with $4.6Nm^{-1}$ substrate stiffness (Fig. B3 in Appendix B). We note that elevated re-clustering rates only indicates the occurrence of effective damping *maxima* and doesn't necessarily correlate to the instantaneous value of effective damping. Transient energy localization in the micro-granular chain is also observed to be robust and be realized over a broad range of parameters (see also Fig. B1). However, due to the highly nonlinear nature of the dynamics it is not possible to predict or control precisely where or when clustering and energy localization occur.

*(Figure 6)*

The elastic substrate stiffness plays an important role in the dynamics despite it being much "softer" compared to the 'stiffer' bead-bead dynamics. The nominal clusters sizes are tunable through the substrate stiffness. Beads within a cluster tend to agglomerate and become compressed due to attractive forces, while the elastic substrate provide restoring forces. This dynamic interplay between the bead-bead and bead-substrate interactions tunes the maximum allowable cluster size, and, as a result, a higher number of relatively smaller clusters due to stiffer substrate enhances localization as is evidenced by the abundance of localization for stiffer substrates in Fig. B1. Notably, in macro-scale granular materials, no such clustering is possible; energy localization in one-dimensional granular chains can only be produced by intruders of variable diameters [30,31] or by granular containers [32].

## VI.  CORRELATION BETWEEN RE-CLUSTERING RATE AND EFFECTIVE DAMPING

To study the relation between the re-clustering rate and the effective damping measure for a wide range of substrate stiffness and impulse intensities we performed a series of simulations. For each case we consider a single mean effective damping measure and a single re-clustering rate. Discretized values of effective damping are defined as the mean effective damping for time periods between re-clustering events (see Appendix A for more details). As discussed previously, for each simulated case we only consider the time period with maxima in effective damping and choose the maximum discretized effective damping segment to be representative of the whole test case. A re-clustering rate for this segment is estimated as the difference in number of clusters in the segment and the following segment divided by the lifespan of the segment. Fig. 7 depicts the maximum discretized effective damping plotted against the log of the corresponding discretized re-clustering rate for 91 test cases over the same range of parameters shown in Figs. B1-B3. The impulse intensity and substrate stiffness for each case are also depicted through the size and color of each 'data circle' on Fig. 7, respectively. This allows us to establish the presence of a positive correlation between effective damping and re-clustering rate over the range of parameters tested. No clear trend between substrate stiffness and impulse energy in relation to effective damping is observed. At this point, we can't make conclusions on the significance of the slope value as it varies according to the number of data points. We note however the existence of a consistently positive slope for different sized data sets, which implies that there is, in fact, a significant positive correlation between the maximum effective damping and its corresponding re-clustering rate over the range of impulse energies and substrate stiffnesses considered. We further add that this trend is also continued if the second highest effective damping segments are included, however, this introduces several outlying points that are more than two standard deviations away from the line of best fit and can be traced back to emergence of numerical artifacts from the discretization algorithm.

*(__Figure 7__)*

# VII.    CONCLUDING REMARKS

Homogeneous micro-granular chains exhibit fundamentally different dynamics compared to their macro-scale counterparts, and have increased capacity for intrinsic energy dissipation. In addition to Hertzian effects, the attractive nature of the beads at the small scale manifests itself in the form of snapping discontinuities as well as dissipation through hysteresis in the force law; to the authors' knowledge it is the first time where Hertzian and hysteretic adhesive effects are combined in the same granular metamaterial. The Nesterenko solitary pulses propagating with no dispersion in macro-granular homogeneous chains disintegrate in the micro-scale, and new strongly nonlinear phenomena such as transient clustering and low-to high frequency energy transfers emerge. Most importantly, cluster formation and re-formation increase the effective dissipation for a wide range of system stiffness and impulse energies. Clustering in micro-granular chains can be considered as analogous to release of pre-stored potential energy in materials with negative stiffness inclusions due to transitions from local buckling states (or 'phase transformation') of the inclusions, causing similar increase in damping (as measured by $\tan \Delta$) in these materials [26-29]. In this study, however, this potential energy release occurs in a transient dynamic context, and in a metamaterial system that combines Hertz-like granular interactions and hysteretic adhesive effects. The new phenomena discussed are shown to be robust and are predicted to exist over a wide range of physically feasible conditions. Furthermore, peaks in effective damping are proven to be related to high re-clustering rates on, both, an instantaneous case-by-case basis as well as a more universal correlation over a range of impulse energies and substrate stiffnesses.

## APPENDIX A: SMOOTHING AND DISCRETIZATION

### A. Smoothing of cumulative number of clusters

The cumulative number of clusters is defined as the number of clusters (not necessarily unique) that have existed up to a given time $t$. The cumulative number of clusters is inherently non-smooth as there are periods over which no re-clustering occurs [See Fig. A1(a)]. A smooth equivalent curve is derived by first performing a cubic spline interpolation between the midpoints of each segment. Special treatment of the first and last segments is applied to constrain the starting point to the origin and the spline to be a flat straight line over the final stable segment. However, this treatment alone does not guarantee a sufficiently smooth curve due to numerical artifacts emerging from segments that occur over infinitesimally small times. An additional cubic spline averaged fit is performed to smooth out these artifacts and the final curve is the one that accurately captures the behavior of the non-smooth curve [Fig. A1(a)]. Taking the numerical time derivative of this curve produces the instantaneous re-clustering rate depicted in Fig. 5.

*(Figure A1)*

### B. Discretization of effective damping and re-clustering rates

From the plot of Fig. A1(a) it is clear that the time interval of the simulation can be divided into segments over which no re-clustering events occur. The mean of the instantaneous effective damping $C_{eff}(t)$ within each of these segments is used as a representation of the effective damping for the lifespan of that segment [Fig. A1(b)]. Unlike the discretization of effective damping, discretized re-clustering rate of a segment is simply estimated as the difference in number of clusters between that segment and the next one divided by the segment lifespan. This is schematically depicted for the *m-th* segment in Fig. A1(c), where $C_{eff,m}$ is the corresponding discretized effective damping, and $r_{c,m}$ the discretized re-clustering rate. Although there is physically no re-clustering in each one of the segments, this estimate is still representative of the local re-clustering rate in the neighborhood of the segment and is qualitatively similar to the smooth curve in Fig. A1(a) such that clusters with longer lifespans lower the local re-clustering rate and the simultaneous formation of many clusters raise the local re-clustering rate.

By inspection, maxima in effective damping coincide with steepening of the slope of cumulative number of clusters, therefore we are only concerned with maxima of the discretized effective damping curve and we choose the segment with the maximum discretized effective damping and its corresponding discretized re-clustering rate to represent its respective case as a single data point (circle) in Fig. 7.

## APPENDIX B: PARAMETRIC STUDIES

***(Figures B1, B2, B3)***

**FIGURES**

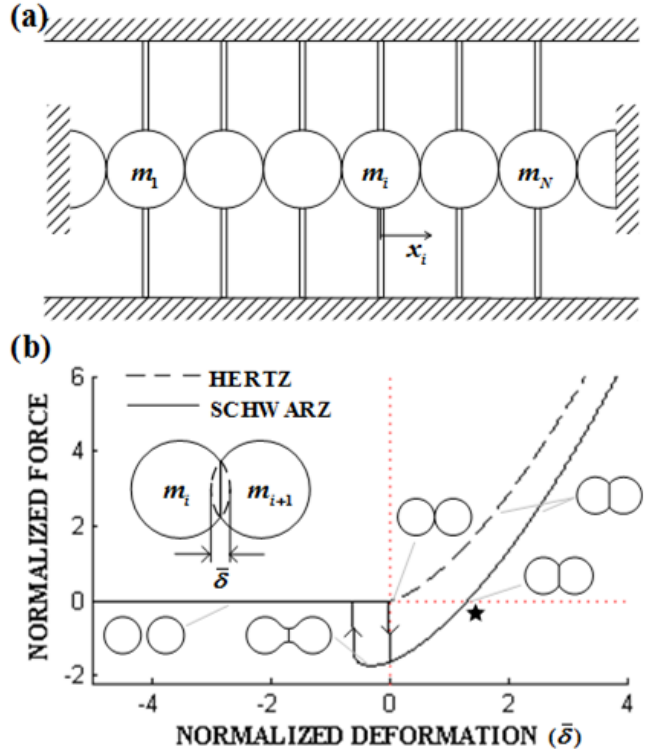

FIG. 1. (a) Spring supported homogeneous micro-granular chain bound by fixed end beads; (b) force-deformation relationship for the Hertz contact model (macro-scale) based on a 3/2 power law (dashed line) and corresponding relationship for the Schwarz contact model (micro-scale) indicating zero force for large separations, a hysteretic loop due to snap-to-contact and snap-from-contact interactions at intermediate separations, and Hertz-like force in compression (solid line); Inset diagrams represent the states of neighboring bead deformations at various points along the force curves.

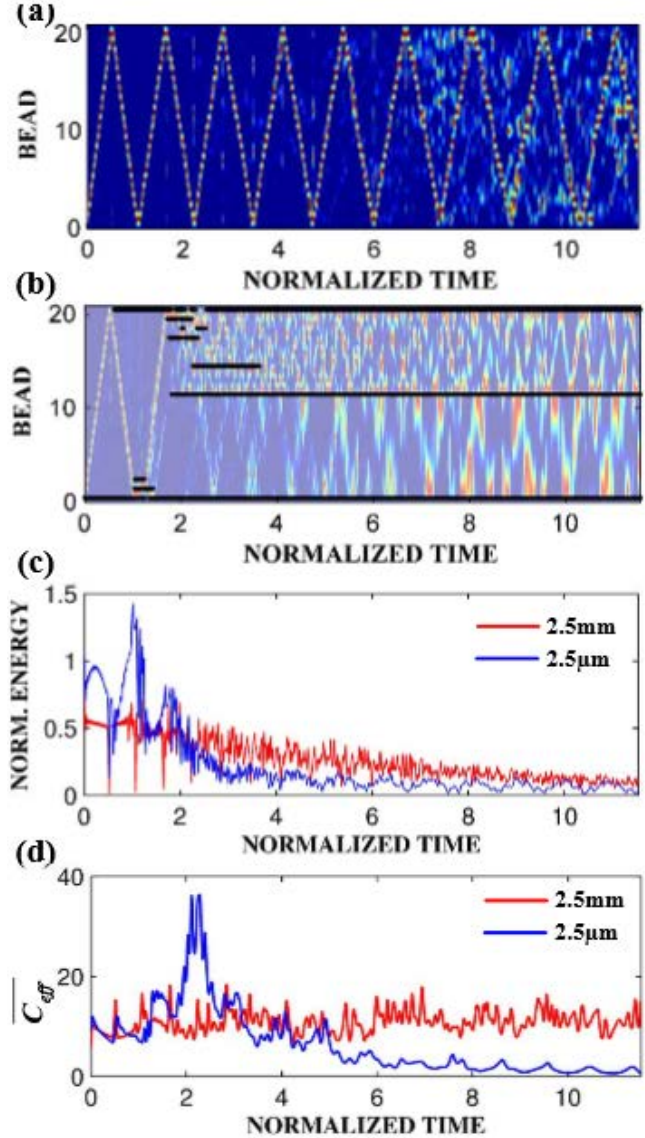

FIG. 2. (Color online) Spatiotemporal evolution of kinetic energy (red is high, blue is low) in, (a) the macro-granular chain of $2.5mm$ particle diameter with solitary pulse propagation; (b) the micro-granular chain of $2.5\mu m$ particle diameter with pulse disintegration and emergence of clustering; (c) evolution of the normalized instantaneous kinetic energies of the media showing enhanced dissipation in the micro-scale due to adhesive effects; (d) normalized effective damping coefficient $\overline{C}_{eff}(t)$ plotted against normalized time for the macro- and micro-granular chains; for the micro-granular chain increases in damping ratio correlate with cluster formation in (b).

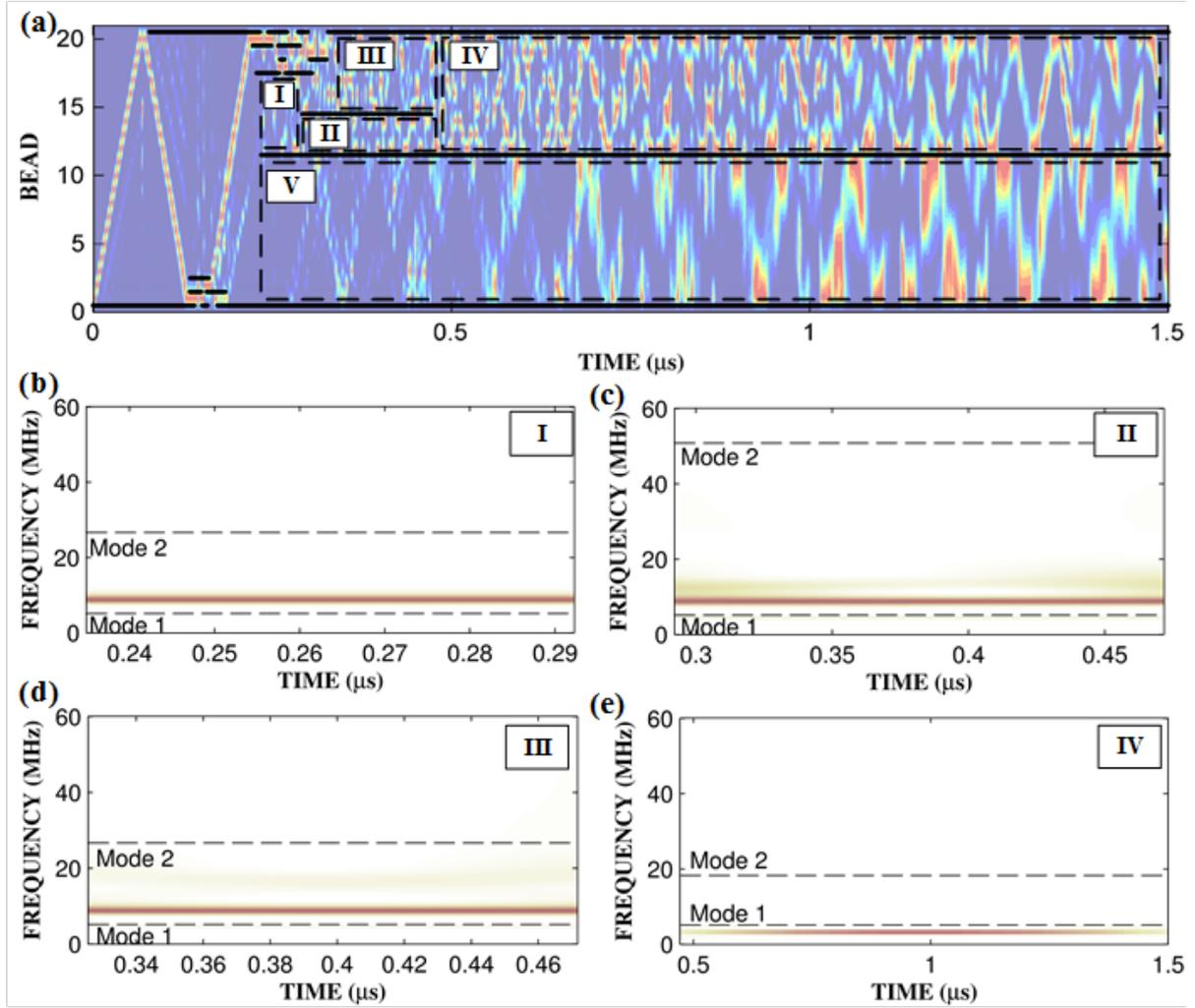

FIG. 3. (Color online) Clustering in the micro-granular chain considered in Fig. 2: (a) Spatiotemporal distribution of kinetic energy (red is more, blue is less) with solid line segments showing cluster boundaries and dashed boxes denoting clusters I–V; (b-e) wavelet transform spectra of the effective particles in the clusters in regions I–IV. Dashed lines denote first and second linearized modes of the clusters. Wavelet transform spectra of the responses in cluster V (not shown) are qualitatively similar to those of cluster IV.

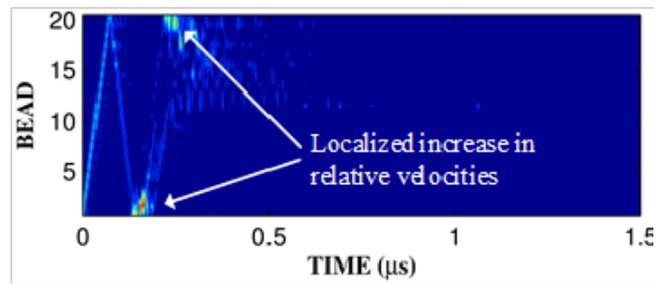

FIG 4. (Color online) Spatiotemporal evolution of the amplitude of relative velocities (red is high) between neighboring beads in the micro-granular chain considered in Figs. 2 and 3; note that, unlike Figs 2(a, b) and Fig. 3(a), in this plot there is no normalization of the relative velocity amplitudes at each time step.

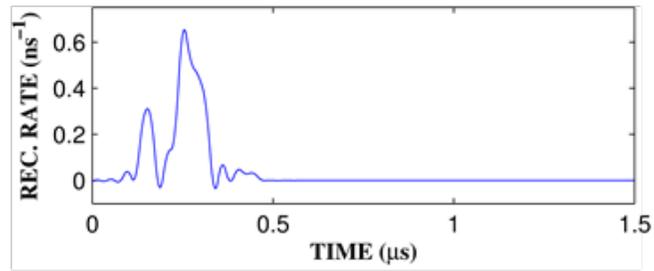

FIG 5. Re-clustering rate plotted against time for the micro-granular chain depicted in Fig. 2 (a, c, d); increases in re-clustering rate are observed to coincide with increases in effective damping coefficient $C_{eff}(t)$ in Fig. 2(d).

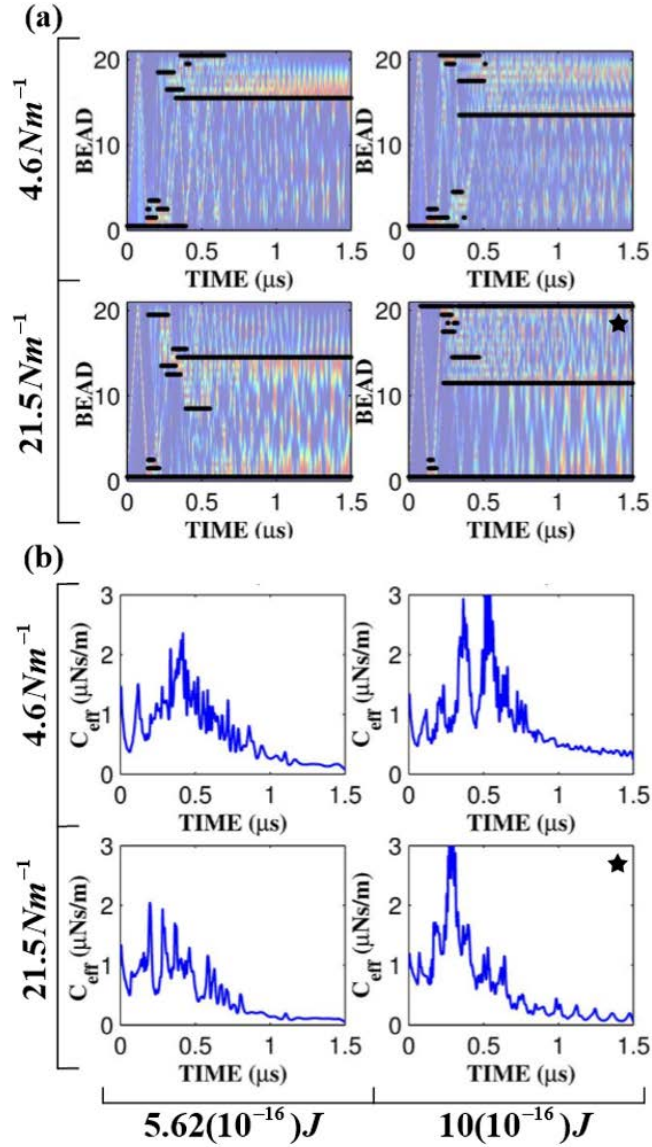

FIG 6. (Color online) Excerpt of parametric study of the impulsively excited 20 bead micro-granular chain (see Appendix B for more cases) for varying substrate stiffness (left scale) and impulse energies (bottom scale): (a) Spatiotemporal distribution of kinetic energy (b) effective damping coefficient corresponding to (a); case denoted by star is the one considered in Figs. 2 and 3.

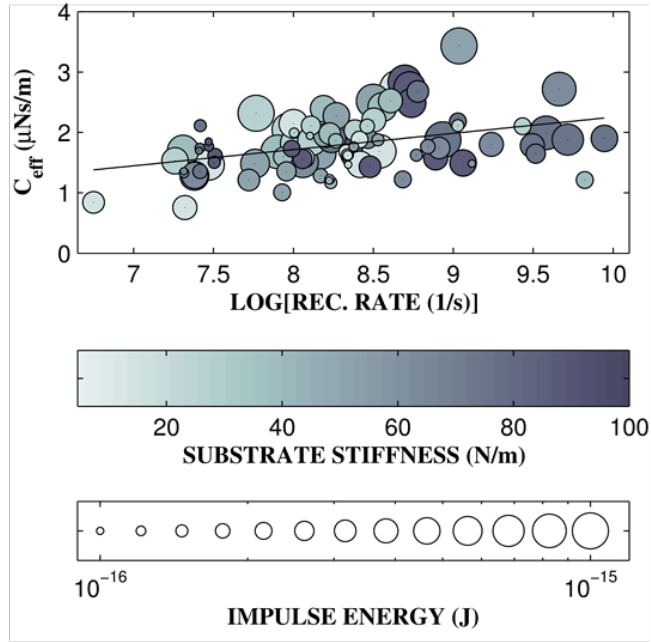

FIG 7. (Color Online) Maximum effective damping plotted against the log of the re-clustering rate for the corresponding segment [see Appendix A and Fig. A1(c) for more details] of the test simulations; grayscale of each data circle relates to substrate stiffness, whereas the radius of the data circle relates to impulse intensity — as shown in the scale bars (note log scale of impulse intensity scale bar).

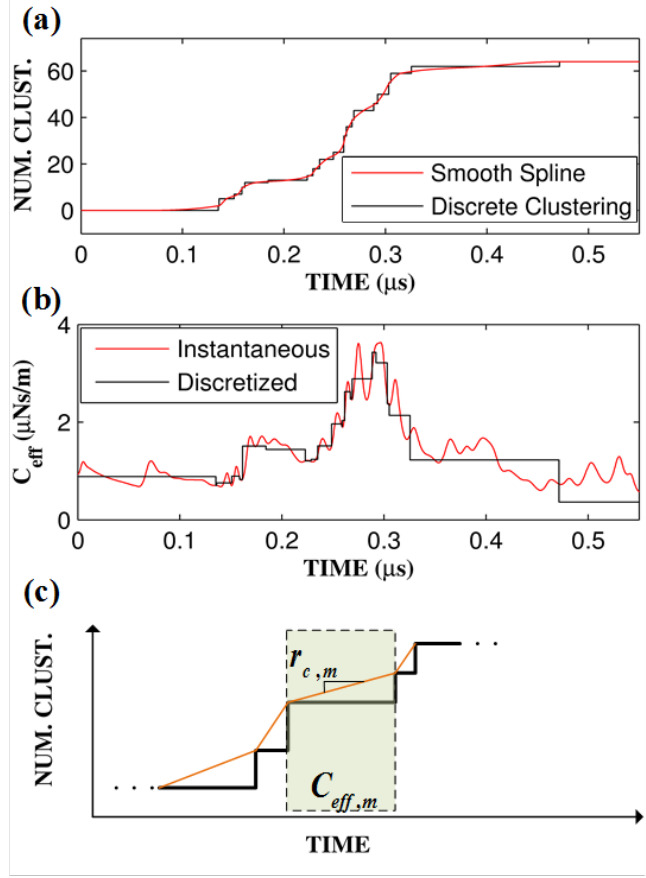

FIG A1. (Color online) (a) Cumulative number of newly formed clusters and smooth spline fit for the same case as in Fig. 2(b); (b) instantaneous effective damping $C_{eff}(t)$ and its corresponding segment averaged discretized effective damping for the same case as in Fig. 2(b); (c) schematic illustrating the definition of discretized re-clustering rate $r_{c,m}$ for the $m$-$th$ segment (highlighted) as the slope of the orange line joining the leading edge of two segments; note the truncated time axis for (a) and (b) since no re-clustering occurs after this time.

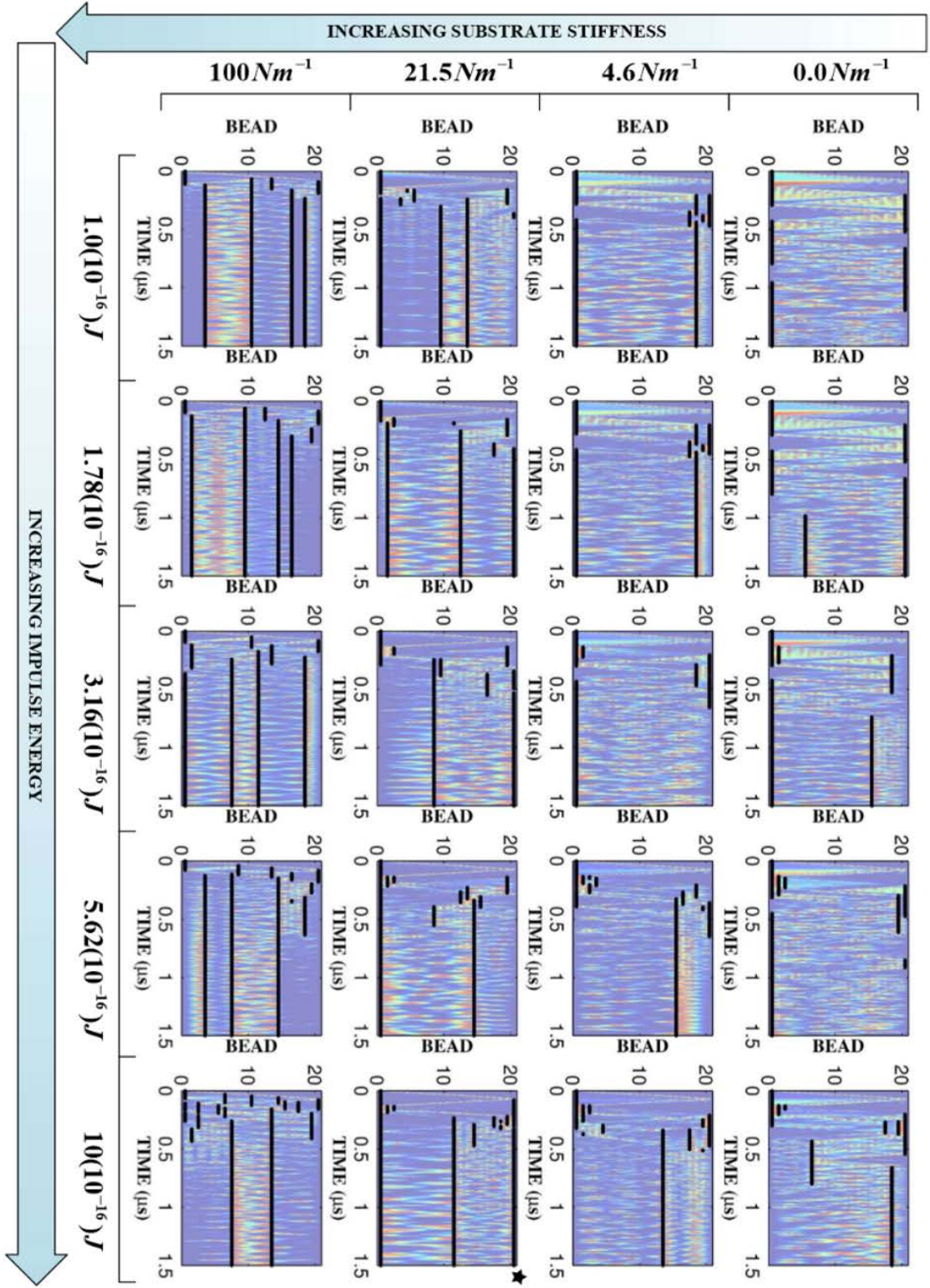

FIG B1. Expanded parametric study of spatiotemporal distribution of kinetic energy of impulsively excited micro-granular chains composed of $N$=20 beads and varying impulse intensity (energy) and substrate stiffness [see also Fig. 6(a)].

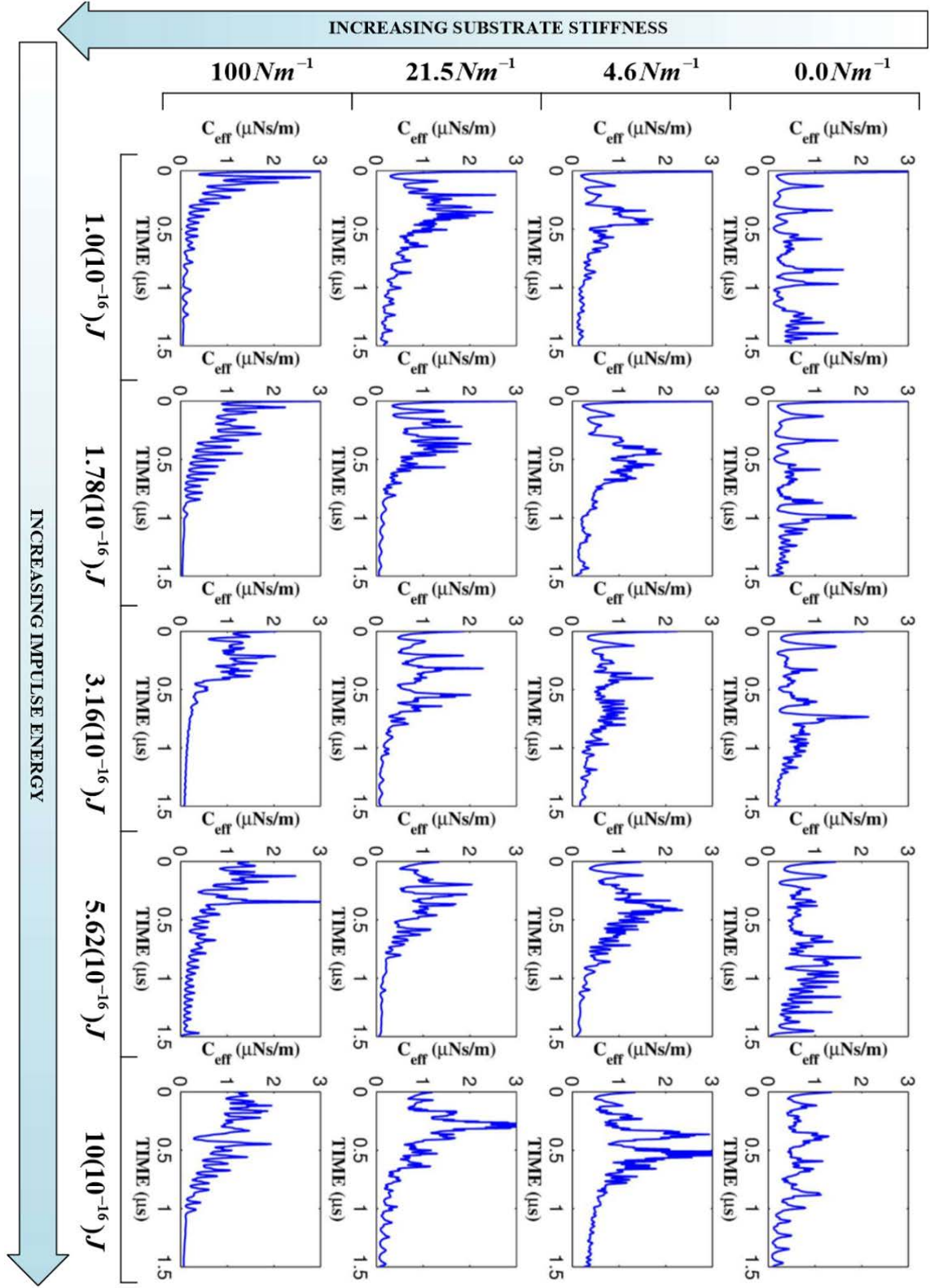

FIG B2. Expanded parametric study of effective damping coefficient corresponding to the plots of Fig. B1 [see also Fig. 6(b)].

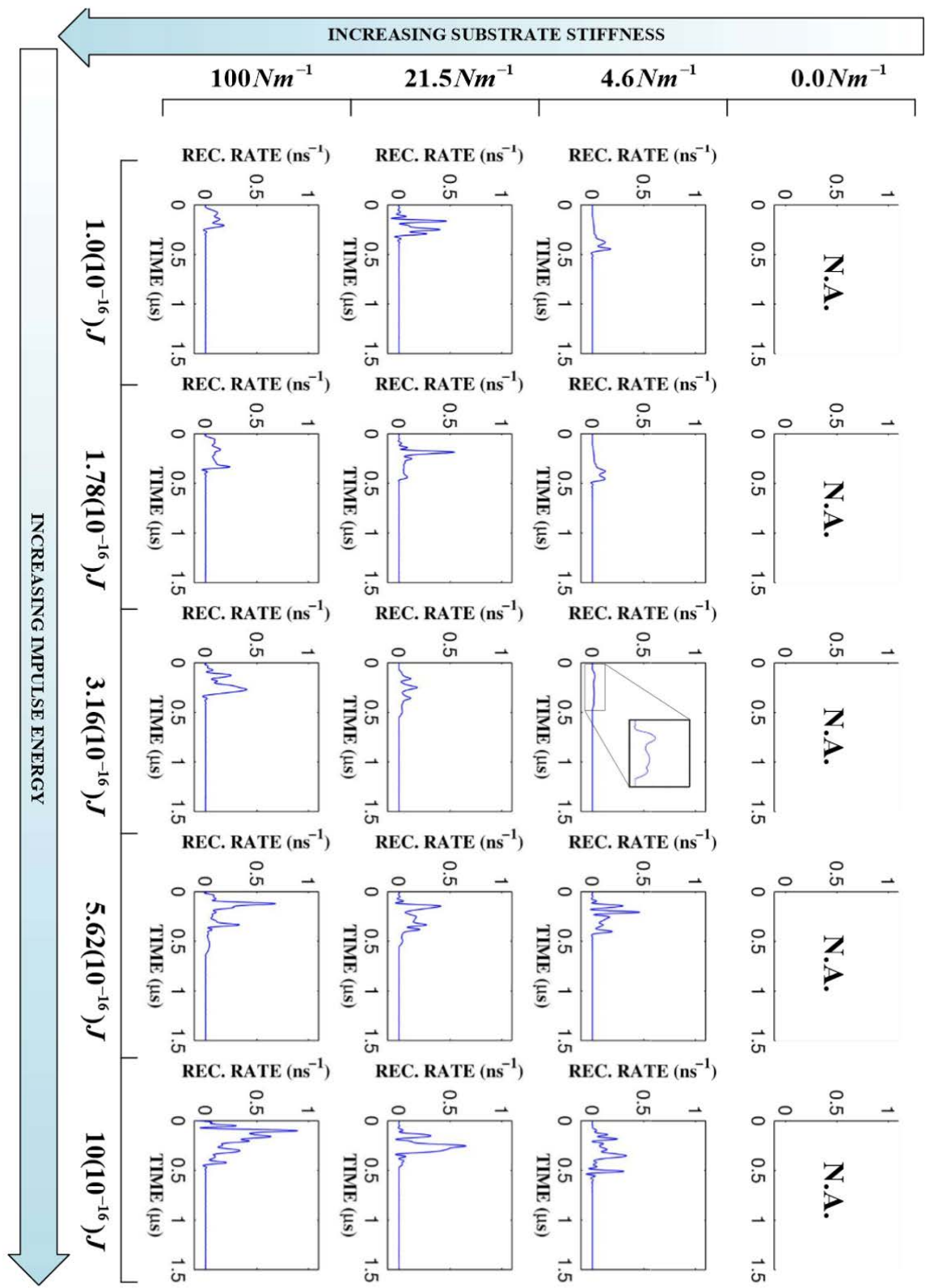

FIG B3. Expanded parametric study of re-clustering rate corresponding to Figs. B1 and B2 [see also Fig.5].